\documentclass[pre,preprint,showpacs,showkeys,preprintnumbers,amsmath,amssymb]{revtex4}
\usepackage{graphicx}   % need for figures
\usepackage{hyperref}   % use for hypertext links, including those to external documents and URLs
\usepackage{dcolumn}% Align table columns on decimal point
\usepackage{bm}% bold math
\begin{document}

\title{Superdiffusion in a Model for Diffusion in a Molecularly Crowded Environment}
\author{Dietrich Stauffer}
\email{stauffer@thp.Uni-Koeln.DE}
\author{Christian Schulze}
\affiliation{Institut f\"ur Theoretische Physik, Universit\"at zu K\"oln,
D--50923 K\"oln, Germany }
\author{Dieter~W.\ Heermann }
\email{heermann@tphys.uni-heidelberg.de}
\affiliation{Institut f\"ur Theoretische Physik, Universit\"at Heidelberg,
Philosophenweg 19, D--69120 Heidelberg, Germany }
\date{\today}
\begin{abstract}
We present a model for diffusion in a molecularly crowded environment. The model
consists of random barriers in percolation network.  Random walks in the presence 
of slowly moving barriers show normal diffusion
for long times, but anomalous diffusion at intermediate times. The effective
exponents for square distance versus time usually are below one at these 
intermediate times, but can be also larger than one for high barrier 
concentrations. Thus we observe sub- as well as super-diffusion in a crowded
environment.
\end{abstract}

\pacs{05.20.Dd, 05.50.+q, 87.16.Ac}% PACS, the Physics and Astronomy
                             % Classification Scheme.

\keywords{anomalous diffusion, effective exponents, random walk}%Use showkeys class option if keyword
                              %display desired
\maketitle

%----------------------------------------------------------------------------------------------
\section{Introduction}
%----------------------------------------------------------------------------------------------

One of the important issues in biology is to understand how diffusion is affected by the environment.
This understanding is needed to correctly describe the passive intracellular transport as this
process my regulate important cellular properties: signal transduction~\cite{Pederson_2000},
gene transcription~\cite{Guthold_1999}, kinetics of reactions~\cite{Berry_2002} and 
regulation of cell polarization~\cite{Valdez-Taubas_2003}.

The interior of biological cells~\cite{Weiss} represents a very dense and 
crowded environment with a specific molecular mobility. Intracellular diffusion is hindered by 
barriers consisting of large molecules sometimes even immobile tethered large molecules, 
binding and collisional interactions. One way of interpreting such a system 
is to view it as a disordered system. In general for random walks~\cite{Binder-Heermann}
in media with disordered  microscopic substructures one expects anomalous 
diffusion~\cite{Lindenberg_2007,Frey_2005,Bouchaud} where 
the mean-square displacement $\langle \Delta{\mathbf r}(t)^2 \rangle$ 
no longer is proportional to the time $t$:

\begin{equation}
\langle \Delta {\mathbf r}(t)^2 \rangle = C_\alpha \; t^\alpha
\end{equation}
with $C_\alpha > 0$.  If $0 < \alpha < 1$ then we call the diffusion 
sub-diffusive, and if $\alpha > 1 $ super-diffusive; normal diffusion has 
$\alpha = 1$.   In biological systems as well as
for models of biological system, so far only sub-diffusive behavior has been 
observed~\cite{Lindenberg_2007} except if the transport is facilitated or restricted. 
For example if the diffusion is directed by a motor protein, non-random super-diffusion can be
observed but not for free diffusion.
However, super-diffusion has been observed in a 
two-dimensional complex plasma~\cite{Ratynskaia_2006}
and in two-dimensional Yukawa liquids~\cite{Liu_2007}. Thus super-diffusion 
does exist and its existence in the cell needs to be discussed.

Sub-diffusion occurs if the mobility of diffusing particle or molecule is impaired by 
obstacles (mobile and immobile) or attractive forces. Under these premises it seems unlikely
to find super-diffusion. 

From a physical point of view the disorder in the cell has a characteristic
time scale. If the diffusing particle has diffused long enough, such that the ti
me
scale has been explored, one expects normal diffusion. If there is no characteri
stic time
(in fractal media) then the diffusion is always anomalous~\cite{Bouchaud}. Howev
er,
this picture does not take into account the (im)mobility, collision, attractive 
forces etc. 
To study the effect produced by immobile as well as mobile barriers we have deve
loped
a model mimicking the molecularly crowded environment with mobile as well as
immobile barriers taking into account particular length scales.

\begin{figure}
\begin{center}
\includegraphics[scale=1.0]{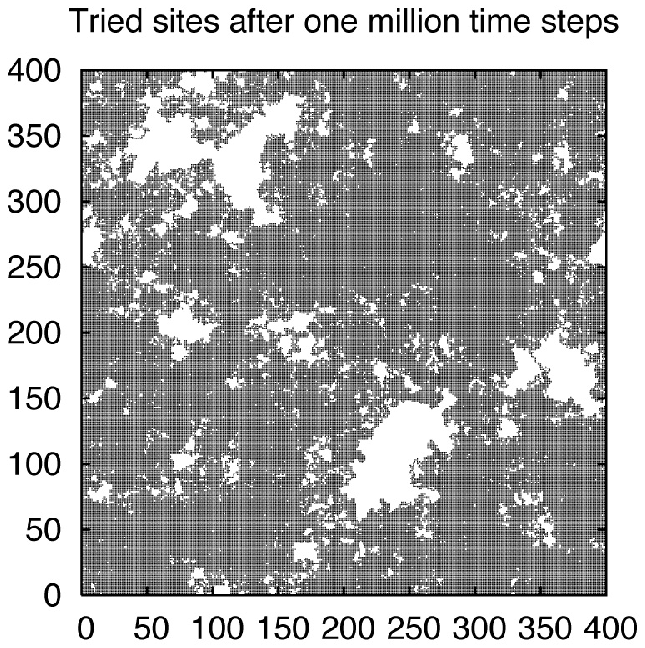}
\includegraphics[scale=1.0]{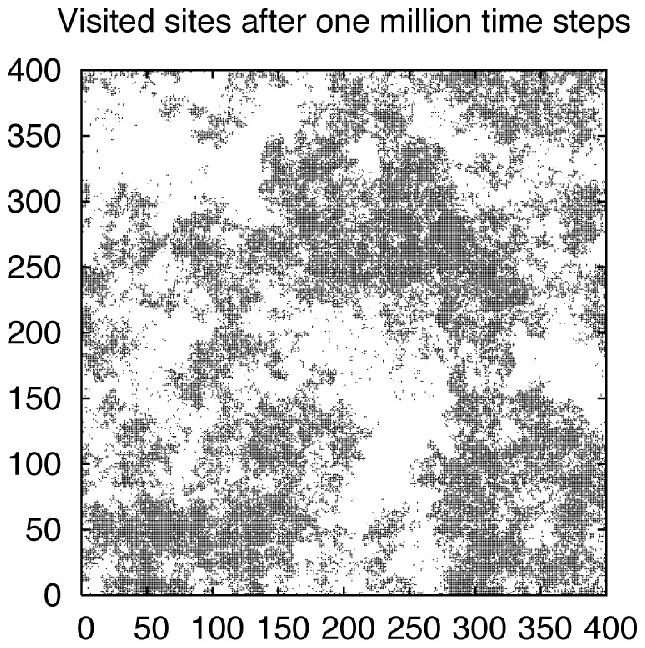}
\caption{\label{fig:stauffer-heermann-1.ps}Illustration in small square lattice:
Tried (part a) and visited (part b) sites after one million time steps.}
\end{center}
\end{figure}

\begin{figure}
\begin{center}
\includegraphics[angle=-90,width=\columnwidth]{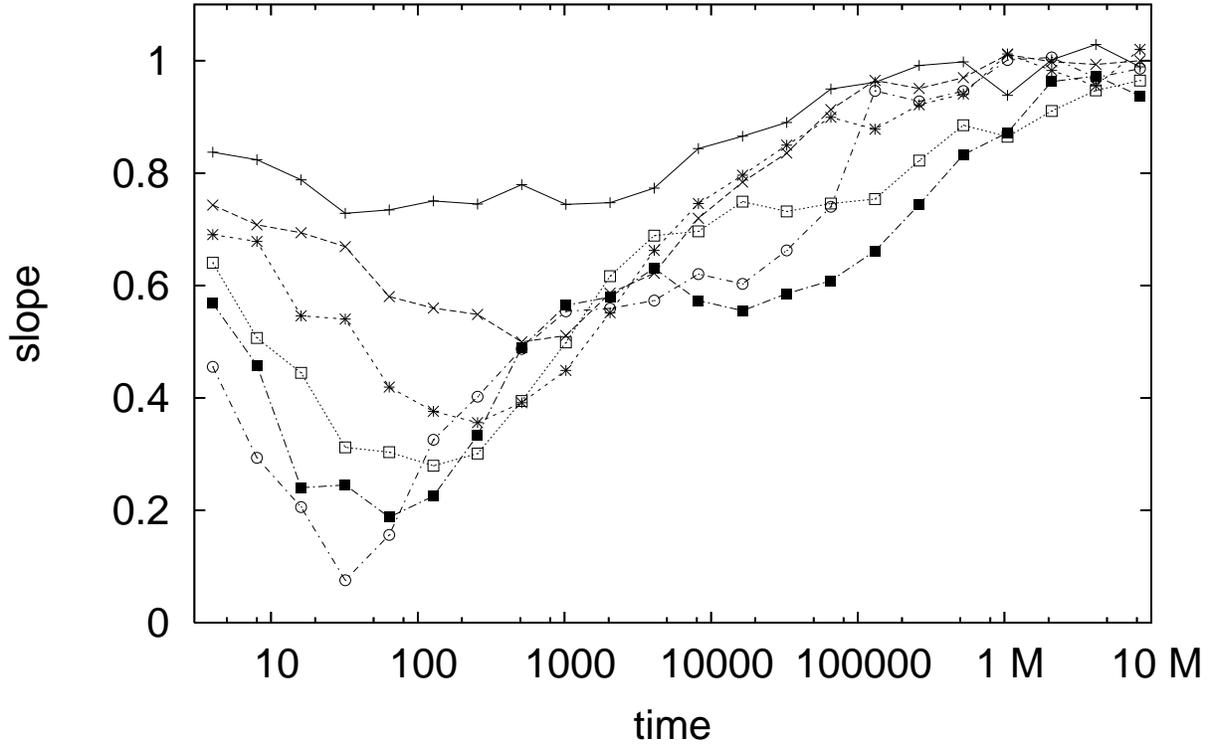}
\caption{\label{fig:stauffer-heermann-2.ps}Effective exponent for 8000 walks on $7001 \times 7001, \; a=1/2,\; 
A=0.01$. The concentration $p$ of allowed sites decreases from top to bottom:
$p_c-p = 0$ (+), 0.1 ($\times$), 0.2 (stars), 0.3 (open squares), 0.4 (full 
squares, 0.5 (circles).}
\end{center}
\end{figure}

\begin{figure}
\begin{center}
\includegraphics[angle=-90,width=\columnwidth]{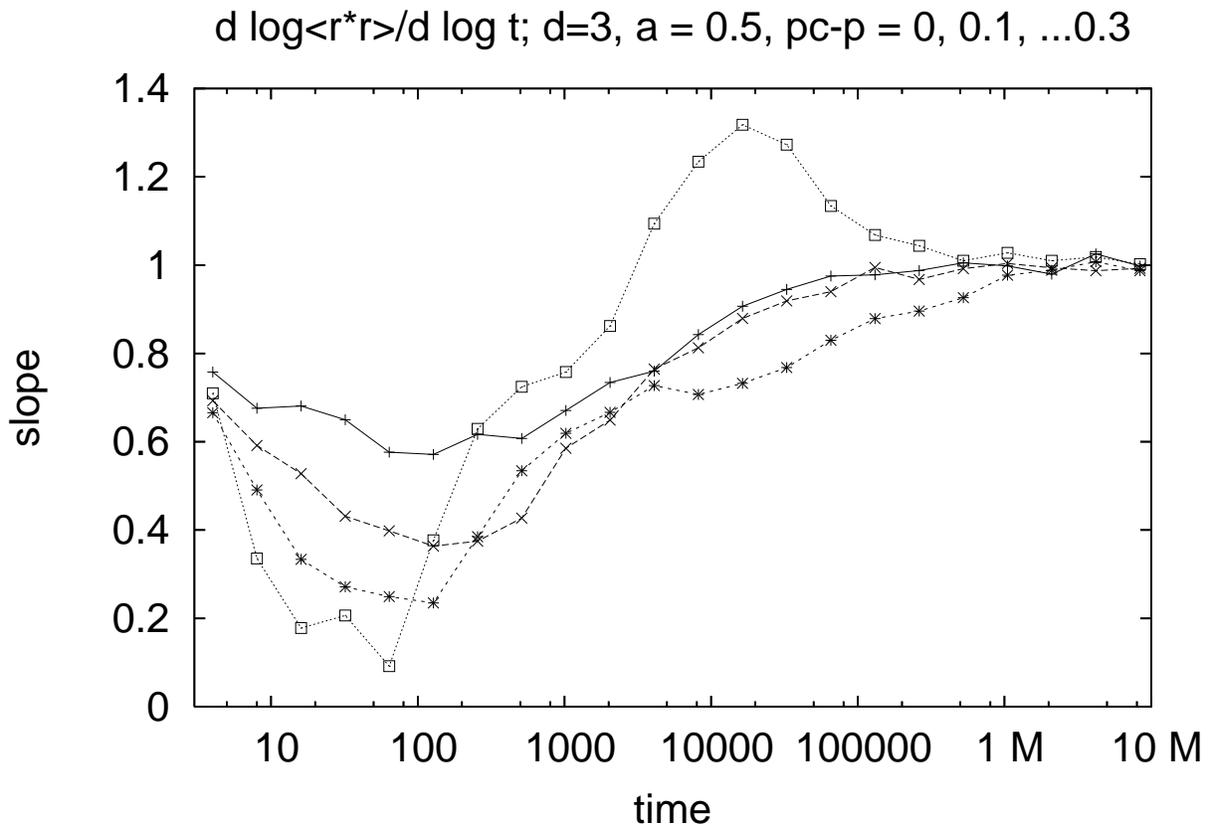}
\includegraphics[angle=-90,width=\columnwidth]{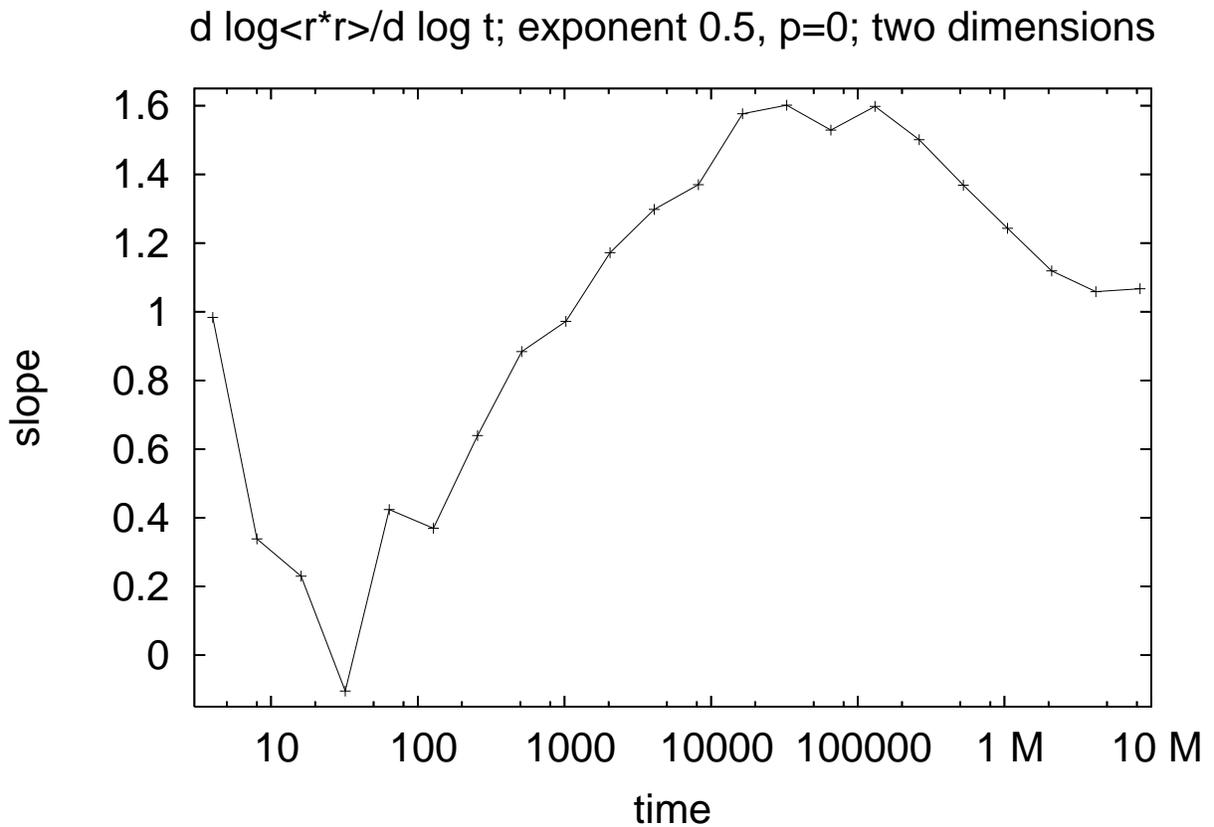}
\caption{\label{fig:stauffer-heermann-3.ps} Effective exponent for 8000 walks on $401 \times 401 \times 401$ (part a, symbols as in Fig.2) and $7001 \times 7001$, (part b); $ a=1/2,\; A=0.01$.}
\end{center}
\end{figure}

\begin{figure}
\begin{center}
\includegraphics[angle=-90,width=\columnwidth]{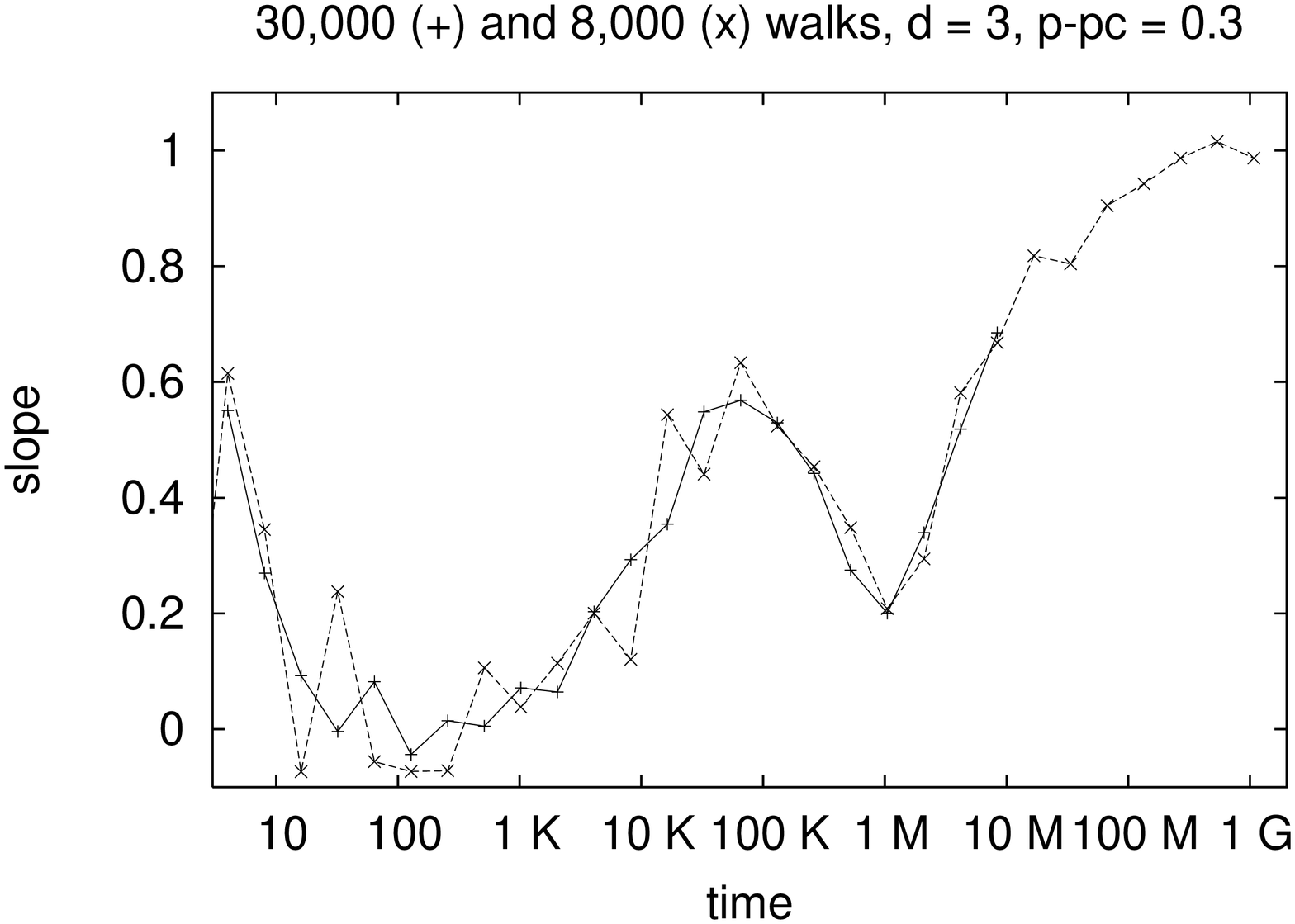}
\caption{\label{fig:stauffer-heermann-4.ps} As Figure \ref{fig:stauffer-heermann-3.ps} but $A = 0.0001$ and better statistics or longer times.}
\end{center}
\end{figure}

%----------------------------------------------------------------------------------------------
\section{The Model}
%----------------------------------------------------------------------------------------------

In percolation theory~\cite{Stauffer-Aharony}, each site on a large lattice is randomly either occupied,
with probability $p$, or empty, with probability $1-p$. For percolative 
diffusion, a random walk starts on an occupied site and then on each time step 
selects randomly a direction to move to. It actually moves one unit distance 
(lattice constant) into this direction
if that neighbor site is occupied. Empty sites are prohibited for the
walker. For $p < p_c$ the walk can extend only over the finite cluster
in which it started, while for $p > p_c$  if can diffuse towards infinity
if it started on the infinite cluster. Right at $p = p_c$ anomalous diffusion 
takes place with a mean square distance increasing to infinity but with an 
exponent $\alpha < 1$ \cite{Stauffer-Aharony}.

In biological applications, the prohibited sites may be more or less mobile
biomolecules. Their effect can be taken into account approximately by assuming
that also a prohibited site $i$ allows the walker to move through, with
some probability $q_i$. The reciprocal probability $1/q_i$ then can be 
interpreted as the lifetime of the barrier, in the sense that about once during
that lifetime the barrier moves away for one time step before returning to that
site. Thus we have still a quenched disorder; with annealed disorder where all
lifetimes are the same, we have normal diffusion, squared distance proportional 
to time, with a diffusivity reduced by the (slowly) moving barriers. We now 
assume that the probability distribution function $f(q)$ for the $q_i$ is a 
power law, $$ f(q) \propto 1/q^a \eqno (2) $$
with some exponent $a$ between zero and infinity. More quantitatively, for each
prohibited site we determine, when it is visited for the first time by the 
walker, a random number $r$, homogeneously distributed between 0 and 1, and
then fix $q_i$ for that site $i$ as $$ q_i = A r^{1/(1-a)} \eqno (3) $$
with some free parameter $A$.

%----------------------------------------------------------------------------------------------
\section{Results}
%----------------------------------------------------------------------------------------------

Figure~\ref{fig:stauffer-heermann-1.ps} illustrates for two dimensions at 
$p=p_c-0.5=0.0927$ and exponent $a=-1/2$ the results of one walk after 
one million time steps. Part a shows the set
of sites which have been tried at least once, and part b shows those sites which
have actually been visited inspite of the barriers. After 8 million steps, all 
sites were tried, and after 64 million steps, all sites were visited.
One can get anywhere, provided one has enough time.

For $7001 \times 70001$ square  lattices, where $p_c \simeq 0.593$, our 
simulations show for $A = 0.01$ and 0.0001 that the squared distance
is a complicated function of the time. 
(For $A=1$ and $a > 1$ the $q_i$ are larger than one which makes little sense, 
and for $a=0$ and 0.5 at $A = 1$ the squared distances are close to $t/2$; not 
shown.) We thus look for smaller $A$ at the slopes in the log-log plots, i.e.
at the effective exponents 
$$ \alpha_{\rm eff} = d \ln \langle \Delta {\mathbf r}(t)^2 \rangle /dt .\eqno(4)$$
In each case we simulate $p = p_c, \;
p_c - 0.1, \; p_c - 0.2, \; \dots$ down to $p_c -0.5 \simeq 0.093$. We see for 
short times different slopes in our log-log plots, but for long times the 
effective slopes approach unity: Normal diffusion with mean squared distance 
proportional to time. For $a = 2$, for an exponentially decaying distribution
$f(q)$, and for a Weibull distribution (stretched exponential) the time 
variations of the effective exponents were similar but less pronounced.

Experimentally more relevant are three instead of two dimensions, and some
results are shown in Figure~\ref{fig:stauffer-heermann-3.ps}~a, rather similar 
to two dimensions in Figure~\ref{fig:stauffer-heermann-2.ps}. Now 
$p_c=0.3116$. For very small $p = 0.0116$, squares in 
Figure~\ref{fig:stauffer-heermann-3.ps}, we see an overshooting
with an effective exponent $\alpha_{\rm eff}$ above unity at intermediate times;
this is not a statistical fluctuation and shows up in all 20 simulated
samples (not shown). It also was seen in two dimensions at very low $p$, 
Figure~\ref{fig:stauffer-heermann-3.ps}~b.
One may call this effect superdiffusion since for more
than one order of magnitude the exponent is above unity.

Basically, the positive probability of each barrier to move away 
and to let through the random walker means that for sufficiently long times
we always get normal diffusion, $\alpha = 1$. For times which are not long
enough to see this moving-away of the barriers, but long enough for the walker
to explore the whole finite cluster for $p < p_c$ on which it started,
we have $\alpha = 0$. For our moderately small $A = 0.01$ these different 
regimes cannot be reliably separated; that works better for much less
mobile barriers: $A = 0.0001$ in Figure~\ref{fig:stauffer-heermann-4.ps}. There the effective exponents are
about zero for $t \sim 10^2$, show a maximum but no longer overshooting 
below $t \sim 10^5$, and approach unity above $t \sim 10^8$.

%----------------------------------------------------------------------------------------------
\section{Discussion}
%----------------------------------------------------------------------------------------------

In summary, we see a non-monotonic variation of the effective exponents with 
time, showing both subdiffusive and superdiffusive behavior. Asymptotically,
however, the exponent always seems to approach unity for $t \rightarrow \infty$.
In experiments with more limited variation of times, this variation of 
$\alpha_{\rm eff}$ with time could wrongly be interpreted as asymptotic 
subdiffusion or asymptotic superdiffusion; long times \cite{Lindenberg} are 
needed.

%------------------------------------------------------------------------------

\begin{acknowledgments}
One of us (DWH) thanks M. Weiss for the discussions.
\end{acknowledgments}

%------------------------------------------------------------------------------

%%%%%%%%%%%%%%%%%%%%%%%%%%%%%%%%%%%%%%%%%%%%%%%%%%%%%

% bibliography

%%%%%%%%%%%%%%%%%%%%%%%%%%%%%%%%%%%%%%%%%%%%%%%%%%%%%

\end{document}